\documentclass[10pt]{article}

\usepackage{amsmath}
\usepackage{amssymb}

\usepackage{graphicx}

\usepackage{cite}

\usepackage{color} 

\usepackage[labelfont=bf,labelsep=period,justification=raggedright]{caption}

\makeatletter
\renewcommand{\@biblabel}[1]{\quad#1.}
\makeatother

\date{}

\usepackage{amsfonts}
\usepackage{amsmath, amsthm, amssymb} 

\begin{document}

\begin{flushleft}
{\Large
\textbf{Human ovarian reserve from conception to the menopause}
}
\\
W. Hamish B. Wallace$^{1,\ast}$, 
Thomas W. Kelsey$^{2}$ 
\\
\bf{1}  Division of Child Life and Health, Department of Reproductive and Developmental Sciences, University of Edinburgh, Scotland, UK
\\
\bf{2} School of Computer Science, University of St. Andrews, Scotland, UK
\\
$\ast$ E-mail:  hamish.wallace@nhs.net
\end{flushleft}

\section*{Abstract}
The human ovary contains a fixed number of non-growing follicles (NGFs) established before birth that decline with increasing age culminating in the menopause at 50-51 years. The objective of this study is to model the age-related population of NGFs in the human ovary from conception to menopause. 
Data were taken from eight separate quantitative histological studies (n = 325) in which NGF populations at known ages from seven weeks post conception to 51 years (median 32 years) were calculated. The data set was fitted to 20 peak function models, with the results ranked by obtained $r^2$ correlation coefficient. The highest ranked model was chosen.
 Our model matches the log-adjusted NGF population from conception to menopause to a five-parameter asymmetric double Gaussian cumulative (ADC) curve ($r^2$ = 0.81). When restricted to ages up to 25 years, the ADC curve has $r^2$ = 0.95. We estimate that for 95\% of women by the age of 30 years only 12\% of their maximum pre-birth NGF population is present and by the age of 40 years only 3\% remains. Furthermore, we found that the rate of NGF recruitment towards maturation for most women increases from birth until approximately age 14 years then decreases towards the menopause. To our knowledge, this is the first model of ovarian reserve from conception to menopause. This model allows us to estimate the number of NGFs present in the ovary at any given age, suggests that 81\% of the variance in NGF populations is due to age alone, and shows for the first time, to our knowledge, that the rate of NGF recruitment increases from birth to age 14 years then declines with age until menopause. An increased understanding of the dynamics of human ovarian reserve will provide a more scientific basis for fertility counselling for both healthy women and those who have survived gonadotoxic cancer treatments.

\section*{Introduction}
Our current understanding of human ovarian reserve presumes that the ovary establishes several million non growing follicles (NGFs) at around five months of gestational age which is followed by a decline to the menopause when approximately 1,000 remain at an average age of 50-51 years \cite{faddy92,fg}. With approximately 450 ovulatory monthly cycles in the normal human reproductive lifespan, this progressive decline in NGF numbers is attributed to follicle death by apoptosis. 

A number of recent reports have challenged this long held understanding of mammalian reproductive biology by reporting the presence of mitotically-active germ stem cells in juvenile and adult mouse ovaries \cite{johnson2004,johnson2005,zou}. While the presence of germ stem cells within the mammalian ovary that are capable of neo-oogenesis remains controversial \cite{telfer,tilly}, a better understanding of the establishment and decline of the NGF population will be important in determining if neo-oogenesis occurs as part of normal human physiological ageing.

Several studies have reported the number of NGFs at different ages in humans \cite{block51,block53,baker,gougeon,richardson,hansen,bendsen,forabosco}  and constructed mathematical models of NGF decline \cite{faddy92,fg,hansen}.  Some of these studies have suggested that the instantaneous rate of temporal change increases around the age of 37 years, when approximately 25,000 follicles remain, followed by exhaustion of the NGF pool and menopause 12-14 years later \cite{gougeon,richardson}.   These studies have addressed the decline
from birth of the NGF population, but none has included the crucial establishment
phase from conception to a peak at 18-22 weeks gestation. Our study uses more histological data than any previous study, and includes data from prenatal ovaries, allowing us to
analyse complete models involving both population establishment and decline.
The objective of our study is to identify the most robust mathematical model to describe the establishment and decline of NGFs in the human ovary from conception to menopause. This will allow us to estimate the number of NGFs present at a given age and describe the rate of recruitment of NGFs towards maturation or apoptosis with increasing age.

\section*{Results}

\subsection*{The best fitting peak model}

\begin{figure}[!ht]
\begin{center}
\includegraphics[width=4in]{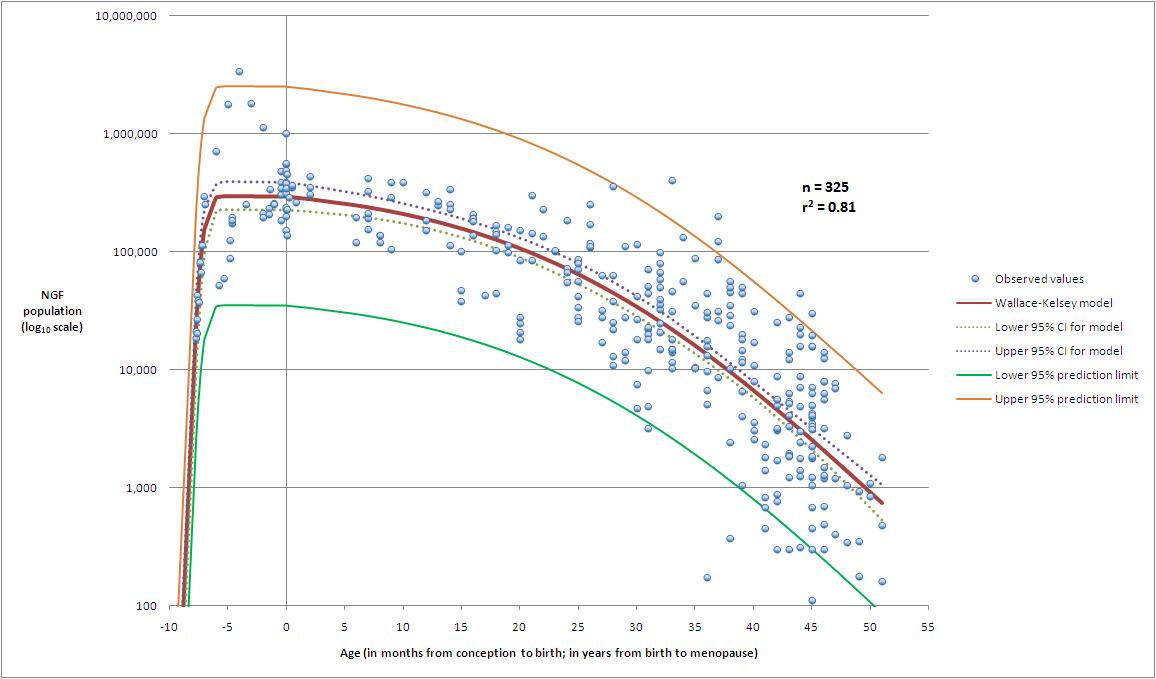}
\end{center}
\caption{
{\bf The model that best fits the histological data.} The best model for the establishment of the NGF population after conception, and the subsequent decline until age at menopause is described by an ADC model with parameters $a$ = 5.56 (95\% CI 5.38-5.74),$b$ = 25.6 (95\% CI 24.9-26.4), $c$ = 52.7 (95\% CI 51.1-54.2), $d$ = 0.074 (95\% CI 0.062-0.085), and $e$ = 24.5 (95\% CI 20.4-28.6).
Our model has correlation coefficient $r^2$ = 0.81, fit standard error = 0.46 and F-value = 364. The figure shows the dataset (n= 325), the model, the 95\% prediction limits of the model, and the 95\% confidence interval for the model. The horizontal axis denotes age in months up to birth at age zero, and age in years from birth to 51 years.}
\label{Fmod}
\end{figure}

  The highest ranked model ($r^2$ = 0.81) returned by the search for the best fitting model to the 325 data-points was a 5-parameter asymmetric double-Gaussian cumulative (ADC) curve (Equation (1)). The first 3 parameters, ($a$, $b$ and $c$) define the scale and amplitude of the curve; the remaining parameters ($d$ and $e$) define the rates of population establishment and decline.
  
\begin{equation}
\qquad \qquad \quad log_{10} (y) = \frac{a}{4} \left[ 1 + \mbox{Erf}\left( \frac{x + b + \frac{c}{2}}{d\sqrt{2}} \right)\right]  \left[ 1 -  \mbox{Erf}\left( \frac{x + b - \frac{c}{2}}{e\sqrt{2}} \right)\right]
\end{equation}

The values for the parameters that maximise the $r^2$ correlation coefficient for our dataset are given in Equation (2).  The model is asymmetric, since rapid establishment is followed by a long period of decline ($d$ is small and $e$ is large); it is double-Gaussian cumulative since it is the product of two Gauss-error functions.

\begin{equation}
log_{10} (NGF) = \frac{5.56}{4} \left[ 1 + \mbox{Erf}\left( \frac{age + 25.6 + \frac{52.7}{2}}{0.074\sqrt{2}} \right)\right]  \left[ 1 -  \mbox{Erf}\left( \frac{age + 25.6 - \frac{52.7}{2}}{24.5\sqrt{2}} \right)\right]
\end{equation}
 
This model (illustrated graphically in Figure 1) demonstrates that 81\% of the variation in individual NGF populations is due to age alone. Full statistical analysis and derivation details for the model are given in additional file 1. Interestingly, if we confine our analysis to the histological data from conception to age 25 years we discover that the ADC model remains the best fit ($r^2$ = 0.95) and that 95\% of the variation in NGF numbers is due to age alone (Figure 2). 
 
 \begin{figure}[!ht]
\begin{center}
\includegraphics[width=4in]{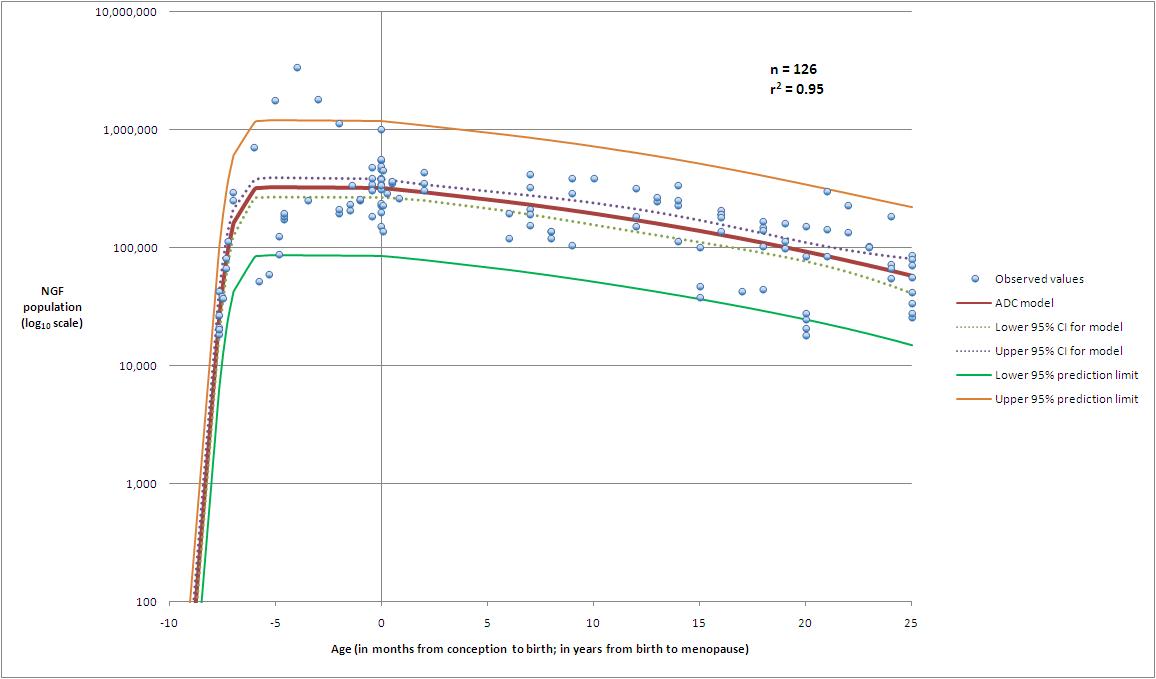}
\end{center}
\caption{
{\bf The model that best fits the histological data for ages up to 25 years.}   The best model for the establishment of the NGF population after conception, and the subsequent decline until 25 years of age is described by an ADC model with parameters $a$ = 5.79 (95\% CI 5.03-6.55), $b$ = 28.0 (95\% CI 15.8-40.2), $c$ = 57.4 (95\% CI 33.1-81.8), $d$ = 0.074 (95\% CI 0.067-0.081), and $e$ = 34.3 (95\% CI -4.2-72.8).
This model has correlation coefficient $r^2$ = 0.95, fit standard error = 0.29 and F-value = 585. This figure shows the dataset (n= 126), the model, the 95\% prediction limits of the model, and the 95\% confidence interval for the model. The horizontal axis denotes age in months up to birth at age zero, and age in years from birth to 25 years.}
\label{Fmod25}
\end{figure}

To guard against model selection bias and to test the robustness of the model with respect to the data, we randomly removed 50 data points 61 times and re-fitted the models, with the ADC model being, on average, the best fitting model (double-sided t-test for difference of means, $p = 0.0065$). Additional file 2 gives details of the statistical tests performed.
  
  \subsection*{Models that allow neo-oogenesis}
  To examine whether a model that permits neo-oogenesis would provide a better fit to the data, we further analysed the data by fitting models that need be neither asymmetric nor single peak. We found that any mathematical model that permits an increase in NGF population after the peak at 18-22 weeks has a markedly inferior fit compared to the best-fitting ADC model (Figure 3).
  
  \begin{figure}[!ht]
\begin{center}
\includegraphics[width=4in]{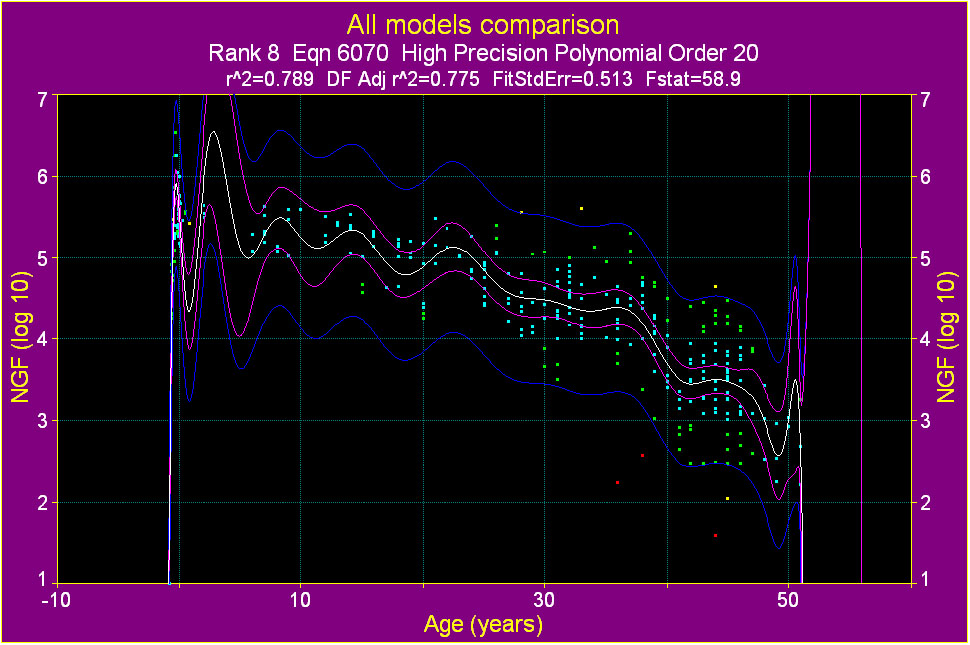}
\end{center}
\caption{
{\bf The highest ranked model that allows growth after the initial peak.}   The highest-ranked non-peak model returned by TableCurve is a polynomial given by $log_{10}(NGF) = c_1*age^{20} + c_2*age^{19} + \cdots + c_{19}*age^2 + c_{20}*age + c_{21}$. Compared to the ADC model for the same data, the model has lower correlation coefficient, higher fit standard error, and lower F-statistic. All other TableCurve models that allow multiple peaks have an inferior fit to the data.}
\label{Fall}
\end{figure}

 \subsection*{Estimated NGF population by age}  
If menopause is defined as a population of less than one thousand (in line with Faddy \& Gosden \cite{fg}), the model predicts age of menopause as 49.6 (95\% CI 47.9 -- 51.2) years, with a 95\% prediction interval of 38.7 -- 60.0 years (Figure 4). Our model gives a maximum mean NGF population  of 300,000 per ovary (95\% CI 225,000 -- 390,000), occurring at 18-22 weeks post-conception, with a 95\% prediction interval (PI) of 35,000 -- 2,534,000 NGFs. Figure 4 gives values for NGF populations at illustrative ages, together with the corresponding 95\% prediction intervals.  Women with an average age of menopause will have around 295,000 NGFs present at birth per ovary, with women destined to have an earlier menopause having around 35,000 NGFs and late menopause women having over 2.5 million NGFs per ovary at birth.   

\begin{figure}[!ht]
\begin{center}
\includegraphics[width=4in]{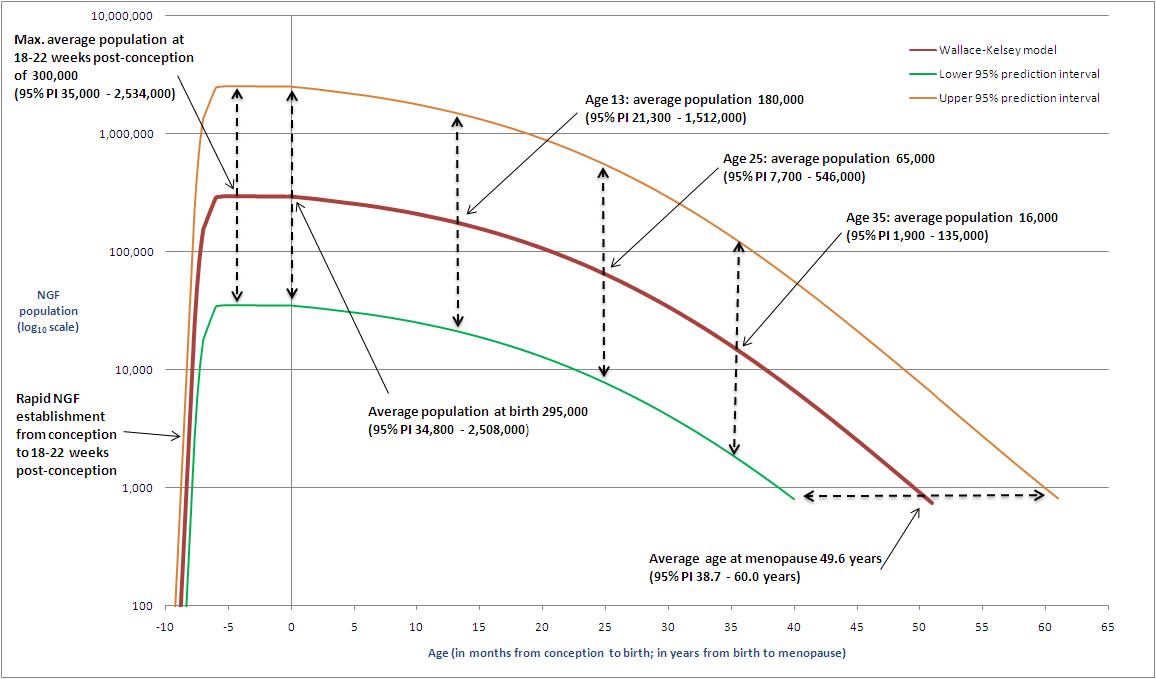}
\end{center}
\caption{
{\bf  Illustrative examples.}   This figure gives illustrative examples of NGF populations predicted by our model. At ages 20 weeks, birth, 13 years, 25 years and 35 years the average NGF population is given, together with the respective 95\% prediction intervals. The predicted average age at menopause (49.6 years) is also shown, together with the 95\% prediction interval.
}
\label{Fex}
\end{figure}

We describe the percentage of the NGF population remaining for a given age for women whose ovarian reserve is established and declines in line with our model (Figure 5).  We estimate that for 95\% of women by the age of 30 years only 12\% of their maximum pre-birth NGF population is present and by the age of 40 years only 3\% remains. The hypothesis that early (respectively late) menopause is related to low (respectively high) peak population at 18-22 weeks post conception is illustrated in  Figure 6.

\begin{figure}[!ht]
\begin{center}
\includegraphics[width=4in]{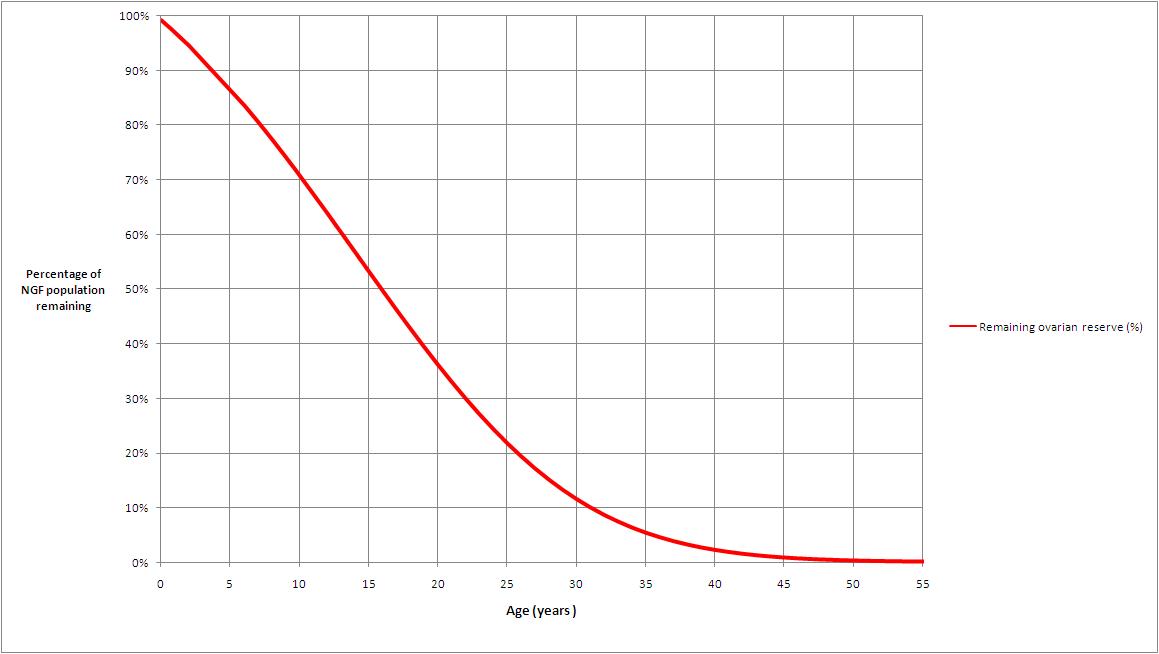}
\end{center}
\caption{
{\bf  Percentage of ovarian reserve related to increasing age.} The curve describes the percentage of ovarian reserve remaining at ages from birth to 55 years, based on the ADC model. 100\% is taken to be the maximum ovarian reserve, occurring at 18-22 weeks post-conception.  The percentages apply to all women whose ovarian reserve declines in line with our model (i.e. late and early menopause are associated with high and low peak NGF populations, respectively). We estimate that for 95\% of women by the age of 30 years only 12\% of their maximum pre-birth NGF population is present and by the age of 40 years only 3\% remains. }
\label{Fpercent}
\end{figure}

\begin{figure}[!ht]
\begin{center}
\includegraphics[width=4in]{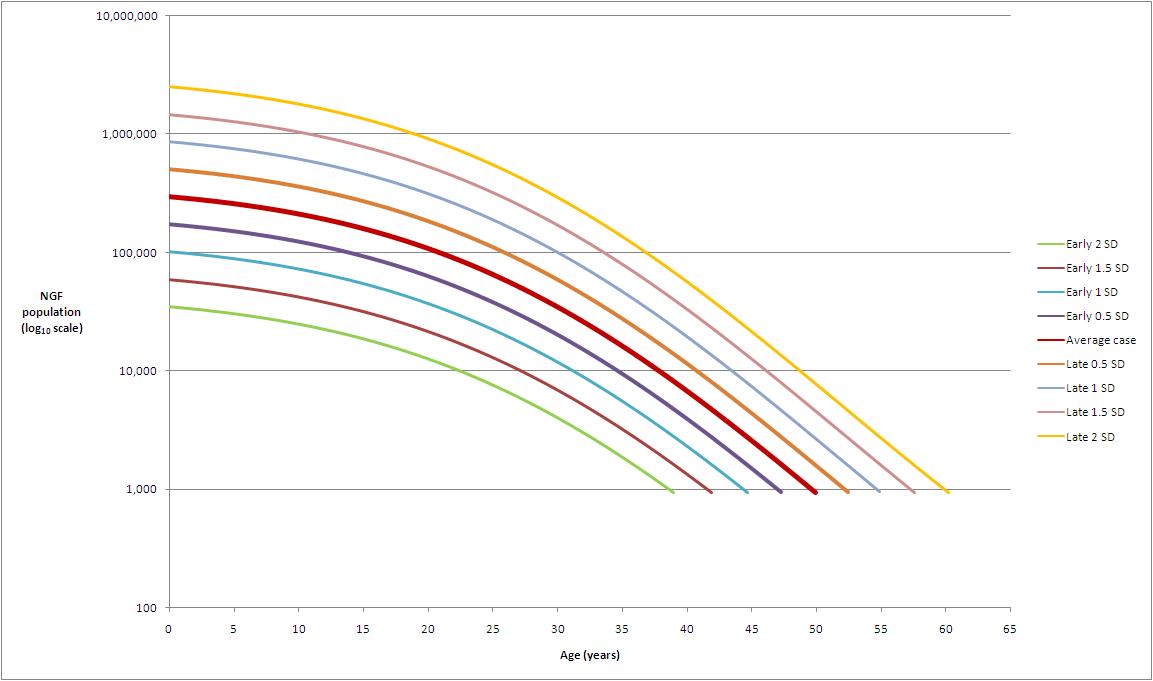}
\end{center}
\caption{
{\bf A hypothetical link between ovarian reserve and age at menopause.}  This figure describes the hypothesis that individual age at menopause is determined by the peak NGF population established at around 20 weeks post-conception. The central curve is the ADC model described in Figures 1 and 4. Above and below are the hypothetical curves for an ovary having log-adjusted peak population varying from the average case by one half, one, one and a half, and two standard deviations. Under this hypothesis, a variation by, for example, one standard deviation in the initial peak population results in a one standard deviation from the average age at menopause. }
\label{Flink}
\end{figure}

 \subsection*{Rates of NGF recruitment towards maturation}
To investigate the number of NGFs recruited towards maturation and ovulation or apoptosis each month we have solved our model to show (Figure 7a) that the maximum recruitment of 880 NGFs per month occurs at 14 years 2 months for the average age at menopause woman. While the maximum rate of recruitment varies hugely, from around 100 NGFs per month (Figure 7b) to over 7,500 NGFs per month (Figure 7c) for women with an early or late menopause respectively, the rate of NGF recruitment increases to a plateau at just over 14 years and then decreases for  women in general irrespective of how many NGFs were established by birth.
  
\begin{figure}[!ht]
\begin{center}
\includegraphics[width=4in]{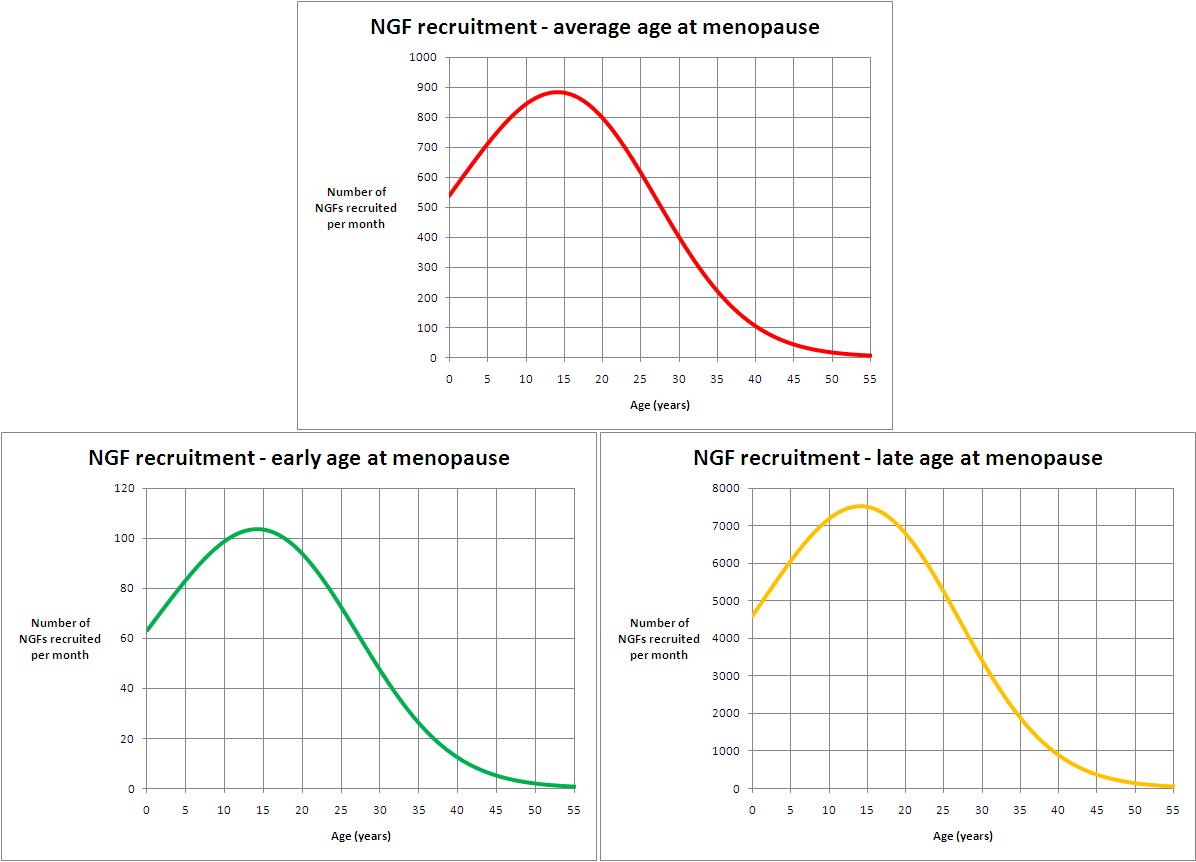}
\end{center}
\caption{
{\bf Rates of NGF recruitment towards maturation.}   Each sub-figure describes the absolute number of NGFs recruited per month, for ages from birth to 55 years, based on population decline predicted by the ADC model.  Figure 7 (a) - red curve - denotes recruitment for individuals whose decline is in line with the average age at menopause; maximum recruitment of 880 follicles per month occurs at 14 years 2 months. Figure 7 (b) - green curve - denotes recruitment for individuals whose decline is in line with early age at menopause (the lower 95\% prediction limit of the model); maximum recruitment of 104 follicles per month occurs at 14 years 2 months. Figure 7 (c) - yellow curve - denote recruitment in line with late age at menopause (the upper 95\% prediction limit of the model); maximum recruitment of 7,520 follicles per month occurs at 14 years 2 months.}
\label{Frates}
\end{figure}

\section*{Discussion}

In this study we have identified the first model of human ovarian reserve from conception to menopause that best fits the combined histological evidence. This model allows us to estimate the number of NGFs present in the ovary at any given age, suggests that 81\% of the variance in NGF populations is due to age alone, and shows that the rate of NGF recruitment increases from birth to age 14 years then declines with age until menopause. Further analysis demonstrated that 95\% of the NGF population variation is due to age alone for ages up to 25 years. The remaining 5\% is due to factors other than age e.g. smoking, BMI, parity and stress. We can speculate that as chronological age increases, factors other than age become more important in determining the rate at which NGFs are lost through apoptosis.

We have made two major assumptions in our study. Firstly, that the results of the eight histological studies that have estimated the total number of NGFs per human ovary are comparable. The definition of a NGF is identical in six of the studies and similar in the remaining two studies. The counting techniques all used a variation of the technique first described by Block \cite{block51}. Our assumption is in line with that of Faddy and Gosden who also assumed histological studies to be comparable when deriving a model for ovarian reserve from birth that also took average age at menopause into account \cite{fg}. The differences between their 1996 study and our study are that we have used more histological data -- including for the first time prenatal data -- and that we use known ranges of age at menopause as a check on the validity of our model, rather than a contributing factor. In the eight reported studies, the majority of younger samples were from autopsy and many of the older subjects had undergone surgical oophorectomy. It is possible that this difference in the source of the ovarian samples influences our finding that  factors other than age become more important in older women. Other studies, and previously reported models, of ovarian reserve have not made a distinction in the reported source of the material; in particular the Hansen et al. study combined  77 autopsy subjects and 45 elective surgical subjects into a single dataset. 

Our second assumption is that the peak number of NGFs at 18-22 weeks gestation defines age at menopause for the individual woman, with early menopause women having low peak populations and late menopause women having high peak populations. The data on the number of NGFs in the ovary is cross-sectional: there is no longitudinal data available and in the absence of a non-invasive test to count NGFs in the individual woman this data is likely to remain unobtainable. Considered together the wide variation at age at menopause and wide variation of peak population of NGFs are suggestive but not conclusive evidence for this assumption to be tenable.

Since the publications by Johnson {\it et al.} \cite{johnson2004,johnson2005} there has been lively scientific debate around the widely held concept that a non-renewing oocyte reserve is laid down in the ovaries at birth, and that neo-oogenesis  does not occur in adult life \cite{telfer}. Johnson and Tilly have argued that their experiments in the adult female mouse have demonstrated conclusively that neo-oogenesis continues in adulthood. They have proposed that the source of postnatal oocyte production is from germline stem cells in the bone marrow, which are transported in the peripheral circulation as germline progenitor cells to arrive in the adult ovary \cite{johnson2005}. The recent report showing isolation and culture of germline stem cells from adult mouse ovaries \cite{zou}, which restored fertility after injection into infertile mice, provides further evidence to support the presence of germ line stem cells in mammalian ovaries. Our analysis of the available histological data demonstrates that any mathematical model that permits an increase in NGF population after the peak at 18-22 weeks has a markedly inferior fit compared to the best-fitting asymmetric peak functions. While the emerging evidence strongly supports the existence of germ stem cells within adult mouse ovaries \cite{tilly}, our model provides no supporting evidence of neo-oogenesis in normal human physiological ageing. 

We have described the percentage of the NGF population remaining for a given age for all women whose ovarian reserve is established and declines in line with our model (Figure 5).  If we assume that a high initial NGF population is associated with late menopause, and that a low peak NGF population is associated with early menopause, then these percentages apply to 95\% of all women.  It is important to note that we have shown that by the age of 30 years the percentage NGF population is already 12\% of the initial reserve and only 3\% of the reserve remains at 40 years of age. A recent study has shown that most women underestimate the extent to which age affects their ability to conceive naturally \cite{bretherick}.

Our finding that the rate of NGF recruitment increases to a plateau at just over 14 years and then decreases in all women irrespective of how many NGFs were established by birth is highly unlikely to be explained by coincidence. From the first comprehensive model of NGF decline from birth \cite{fg} we can calculate that the maximum NGF recruitment occurs at birth (data not shown). However, this model was not only based on goodness of fit to histological data, but it also included adjustments to take known distribution of ages at menopause into account. A more recent model of decline from birth is based entirely on fitting to histological data \cite{hansen}. For this model we calculate that the maximum recruitment of NGFs to maturation occurs at 18 years 11 months (data not shown). 

In western society the average age of menarche is around 13 years \cite{parent}, with early breast development appearing around age 11 years. Our data suggests that the onset of oestrogenisation and ovulation heralds a slowing in the rate of NGF recruitment. Our findings suggest that both endocrine and paracrine factors may be important in the slowing and subsequent decline in the rate of NGF recruitment.  An important candidate is anti-M{\"u}llerian hormone (AMH), a member of the transforming growth factor-beta (TGF-$\beta$) superfamily of growth factors \cite{knight}. They are produced by ovarian granulosa cells and oocytes in a developmental, stage-related manner and function as intra-ovarian regulators of folliculogenesis. There is good evidence that AMH from granulosa cells of pre-antral or antral follicles exerts a negative inhibitory influence on the primordial to primary follicle transition \cite{broekmans}. Furthermore AMH has been proposed as an indirect marker of ovarian reserve in post-pubertal women \cite{broekmans}. Until the onset of puberty (characterised by the switching on of the hypothalamic-pituitary axis and the pulsatile secretion of the gonadotophins FSH and LH) follicular maturation rarely progresses beyond the pre-antral stage. The presence of the pulsatile secretion of FSH and LH at puberty promotes follicular maturation to the antral stage and beyond. There is however incomplete data on AMH levels in pre-pubertal girls in the literature: AMH is undetectable before birth \cite{josso} and is detectable at low levels in infants \cite{guibourdenche}. The explanation for our finding that the rate of NGF recruitment increases until the onset of puberty, levels off at around 14 years of age, and then declines to the menopause remains unclear. It is interesting to speculate that AMH levels which are undetectable at birth may rise at puberty with the establishment of regular ovulatory cycles and be responsible for the slowing of the rate of NGF recruitment that occurs at puberty.
	
Can a more complete understanding of the establishment and decline of the non-renewing pool of NGFs help us to assess ovarian reserve for the individual woman? Several candidate markers for the assessment of ovarian reserve in the individual woman have been suggested including FSH, Inhibin B, AMH, and antral follicle counts and ovarian volume by transvaginal ultrasound \cite{knauff,younis}. We have previously reported a striking correlation between ovarian volume and NGF population using an earlier model \cite{wk2004}. However the measurement of ovarian volume by transvaginal ultrasound is imprecise, particularly at the lower end of the range\cite{brett}. It is likely that a better understanding of NGF establishment and decline will improve our ability to assess ovarian reserve for the individual woman. One immediate application of our model is to better understand the effect of chemotherapy and radiotherapy on the human ovary. Using a model based on less complete histological data, we estimated the radiosensitivity of the human oocyte \cite{wk2003} and were subsequently able to estimate the effective sterilising dose of radiotherapy at a given age for the individual woman \cite{wk2005}. Knowledge of the dose of radiotherapy and age at which it is delivered provides an important opportunity for accurate counselling of women receiving cancer treatment and will help us to predict which women are at high risk of premature menopause and who may therefore benefit from ovarian cryopreservation \cite{wallace}.

  We have described and illustrated a model of human ovarian reserve from conception to menopause that best fits the combined histological evidence. Our model matches the log-adjusted NGF population to a five-parameter asymmetric double Gaussian cumulative (ADC) curve ($r^2$ = 0.81). When restricted to ages below 26 years, the ADC curve has $r^2$ = 0.95. We estimate that for 95\% of women by the age of 30 years only 12\% of their maximum pre-birth NGF population is present and by the age of 40 years only 3\% remains. Furthermore, we found that the rate of NGF recruitment towards maturation for most women increases from birth until approximately age 14 years then decreases towards the menopause. An increased understanding of the dynamics of human ovarian reserve will provide a more scientific basis for fertility counselling for both healthy women and those who have survived gonadotoxic cancer treatments.

\section*{Methods}
 \subsection*{Histological studies of NGF populations}
 
 \begin{table}[!ht]
\caption{
\bf{The eight quantitative histological studies forming the combined dataset}}
\begin{tabular}{|clc|cccc|}
        \hline \multicolumn{3}{|c|}{Study} &  \multicolumn{4}{|c|}{Statistics}\\ 
        Number & First author  & Year & No. ovaries & Min. age & Max. age & Median age\\ \hline \hline
        1 & Bendsen & 2006 & 11 & -0.6 & -0.6 & -0.6 \\ 
         2 & Baker & 1963 & 11 & -0.6 & 7.0 & -0.2 \\ 
         3 & Forabosco & 2007 & 15 & -0.5 & 0.5 & -0.3 \\ 
         4 & Block & 1953 & 19 & -0.2 & 0.0 & 0.0 \\ 
         5 & Hansen & 2008 & 122 & 0.1 & 51.0 & 38.0 \\ 
         6 & Block & 1951 & 86 & 6.0 & 44.0 & 28.0 \\ 
         7 & Gougeon & 1987 & 52 & 25.0 & 46.0 & 39.5 \\ 
         8 & Richardson & 1987 & 9 & 45.0 & 51.0 & 46.0 \\ \hline \cline{4-7}
         \multicolumn{3}{|c|}{Overall} & 325 & -0.6 & 51.0 & 32.0 \\ \hline
      \end{tabular} 
\begin{flushleft}
  All ages are in years; the year column refers to the year of publication. 
\end{flushleft}
\label{Tdata}
 \end{table}

   Data were taken from all eight known published quantitative histological studies of the human ovary (Tables 1 \& 2). In all eight studies the definition of a NGF is similar. For studies 2,4,5,6,7 \& 8 the definition of a NGF is according to the morphological criteria of Block and Gougeon \cite{block51,gougeon}: an NGF was counted when a clearly defined oocyte nucleolus was present within the optical dissector counting frame. In studies 1 \& 3, fetal ovary studies, an NGF was defined as being an oocyte surrounded by a single layer of granulosa cells.  Each of these studies used a variation on the technique developed by Block to estimate the NGF population of an ovary \cite{block51}.  Ovaries are sectioned and some of the sections stained and photographed. A mean NGF volume is calculated from a sample image and used throughout all subsequent calculations. The photographs are analysed by hand, with the number of NGFs appearing in the photograph being counted. By assuming an even distribution throughout the ovary, the population of the samples is integrated into an estimated population for the entire ovary.  The studies differ in the stain used, the number of samples chosen, the method of counting, and the mathematical formula used to obtain the estimated population from the sample populations. Some ovaries were obtained at autopsy and some after elective surgical oophorectomy. Older subjects were more likely to have undergone surgical oophorectomy. No standard error has been calculated for such studies, as the exact number of NGFs in a specific ovary has not been calculated. We combined these data into a single dataset (n=325) and enforced a zero population at conception. 
   
   \begin{table}[!ht]
\caption{
\bf{The eight quantitative histological studies of NGF population summarised by ovarian age}}
\begin{tabular}{|cl|ccccccccccc|}
        \hline \multicolumn{2}{|c}{Study} &  \multicolumn{11}{|c|}{Ovaries per range of ages}\\ 
        No. & First author & $<$ 0 & 0--5 & 5--10 & 10--15 & 15--20 & 20--25 & 25--30 & 30--35 & 35--40 & 40--45 & 45--51 \\ \hline \hline
        1 & Bendsen &  11 & 0 & 0& 0& 0& 0& 0& 0& 0& 0& 0 \\ 
         2 & Baker &   5 & 5 & 1   & 0& 0& 0& 0& 0& 0& 0& 0\\ 
         3 & Forabosco &   14 & 1 & 0& 0& 0& 0& 0& 0& 0& 0& 0 \\ 
         4 & Block &   5 & 14 &  0& 0& 0& 0& 0& 0& 0& 0& 0 \\ 
         5 & Hansen &   0 & 6 & 3 & 4 & 8 & 8 & 8  & 12 & 22 & 26 & 15   \\ 
         6 & Block &  0 & 0 & 10 & 8 & 12 & 8 & 12 & 26 & 10 & 10 & 0 \\ 
         7 & Gougeon &    0 & 0 & 0 & 0 & 0 & 3 & 7 & 9 & 10 & 17 & 6 \\ 
         8 & Richardson &   0 & 0& 0& 0& 0& 0& 0& 0& 0& 2 & 7  \\ \hline \cline{3-13}
         \multicolumn{2}{|c|}{Overall} &  35 & 26 & 14 & 12 & 20 & 19 & 27 & 47 & 42 & 55 & 28 \\ \hline
      \end{tabular}
 \begin{flushleft} Prenatal ovaries are listed in the $<$ 0 column.
\end{flushleft}
\label{Tages}
 \end{table}

  \subsection*{Mathematical models}
 We fitted 20 asymmetric peak models to the data set, using TableCurve-2D (Systat Software Inc., San Jose, California, USA), and ranked by $r^2$ correlation coefficient. Each model defines a generic type of curve and has parameters, which, when instantiated gives a specific curve of that type. For each type of curve, we calculated values for the parameters  that maximise the $r^2$ for that model. The 20 models supplied by TableCurve are those that are commonly reported in the scientific literature as models of datasets that rise and fall  such as pharmacodymanics, cell populations and electromagnetic signals. The Levenberg-Marquardt non-linear curve fitting algorithm was used, with convergence to 15 significant figures in $r^2$ after a maximum of 10,000 iterations. We performed the same analysis on the datapoints associated with an age of 25 years or under. Supporting Information file S1 contains the dataset, the model ranking, the output used to prepare Figures 1 and 2, and the statistics associated with the highest-ranked model.
     
 To avoid selection bias, we randomly removed 50 datapoints 61 times and re-fitted the models, calculating the mean and standard deviations of the $r^2$ coefficients obtained for each model. We then compared the mean $r^2$ obtained for the two highest ranked models for a statistically significant difference. Supporting information file S2 contains statistics regarding the correlation coefficient for the models obtained in this way, together with output for the test for a statistically significant difference of the two highest means. To further avoid selection bias from our initial choice of models, and to allow the possibility of more than one peak (i.e. to allow models involving regeneration of ovarian reserve), we fitted all 266 models supplied by TableCurve, again ranking by $r^2$. The highest ranked model was used as the basis for further calculations.  Under the modelling assumption that, in general, a high (versus low) established population results in a late (versus early) menopause, we calculated the percentage of NGF pool at given ages, and the absolute monthly loss of germ cells from birth until age 55.    

 \subsection*{Calculation of recruitment rates}
 
 To calculate the rates of recruitment of NGFs towards maturation we solved the equations describing the early, late and average menopause models for all months from birth to menopause. The absolute numbers of NGFs recruited were then given by the differences in successive  monthly totals.

\section*{Acknowledgments}
 We would like to acknowledge the critical discussions we have had with Prof Richard Anderson, Prof Ian Gent, Dr Jacob Howe, Dr Dror Meirow and Dr Evelyn Telfer.  

  
\bibliography{WK-PLosOne}

\end{document}